\documentclass[twocolumn,prl,graphics,showpacs,noshowkeys]{revtex4}
\begin{document}
\input{epsf.sty}
\title{A simple statistical explanation for the 
localization of energy in nonlinear lattices\\ 
with two conserved quantities}
\author{Benno Rumpf}
\affiliation{Max-Planck-Institute for the Physics of Complex Systems, \\
N\"othnitzer Stra\ss e 38, 01187 Dresden, Germany
}
\begin{abstract}
The localization of energy in 
the discrete nonlinear Schr\"odinger equation is explained with 
statistical methods. The partition function and the entropy of the system 
are computed for low-amplitude initial conditions. 
Detailed predictions 
for the long-time solution are derived. 
Localized high-amplitude excitations absorb a surplus of energy when they 
emerge as a by-product of the production of entropy in the 
small fluctuations. 
The thermodynamic interpretation of this process 
applies to many dynamical 
systems with two conserved quantities. 
\end{abstract}

\maketitle
Localization of energy within a small number of isolated high-amplitude
structures is a widespread phenomenon in nonlinear optics 
\cite{kel,ace,eisen}, 
plasma physics \cite{zcol}, Bose-Einstein condensates \cite{trom}
and nonlinear lattice dynamics 
\cite{mojo,ras,rune,rune03}. 
Peaks of the energy density result either  
from a collapsing wave train that leads 
to a finite-time blow-up of the amplitude \cite{zcol,zss}, or, 
in spatially discrete systems, from a sequence of merging breathers 
\cite{rune}. 
\\
It is 
the purpose of this article is to show that 
the formation of peaks is driven by the production of entropy and that 
this behavior is generic for the thermalization of many 
conservative systems where a second quantity is conserved in 
addition to the Hamiltonian. 
The discrete nonlinear Schr\"odinger equation is a simple 
generic equation that describes discrete breathers in nonlinear 
optical waveguide arrays \cite{ace,eisen} and dilute 
Bose-Einstein-condensates that are trapped in periodic potentials 
\cite{trom}. 
The spatial discreteness avoids the leakage of energy to infinitesimal 
scales that can occur during the wave-collapse 
of continuous systems \cite{dya} so that  
this system is a simple but representative model for localization 
processes. 
The non-compactness of the phase space leads to a merely technical difficulty 
in the statistical treatment of high-amplitude structures. 
\\ 
Fig.1a shows the focusing process for 
the focusing discrete nonlinear Schr\"odinger equation (DNLS) 
\begin{equation}
\label{dnls}
i\dot{\phi}_n=\phi_{n+1}+\phi_{n-1}+|\phi_{n}|^{2}\phi_{n}
\end{equation}
Any coefficients of this equation can be removed by scaling, and 
consequently all quantities are dimensionless. 
Firstly, breathers with moderate amplitudes appear 
periodically in space and time following a phase instability of a 
regular low-amplitude initial solution. 
Subsequently, they merge into more persistent 
peaks with higher amplitudes (lattice site 188 in Fig.1a).  
The system finally settles into a state 
where immobile high-amplitude peaks 
(the ring with $|\phi|\approx 2.3$ in Fig.1b) 
emerge 
from a low-amplitude disordered background (core with $|\phi|<0.5$). 
The peaks divide the system into patches of the order 
of 100 lattice sites where the amplitude is low and the dynamics 
is irregular. The peaks 
oscillate corresponding to their amplitude-dependent frequency 
and their amplitudes fluctuate slightly, but their position 
in the lattice almost never changes. 
\\
This behavior depends crucially 
on the system's energy $E=<\cal{H}>$, where $
{\cal{H}}=
\sum_i (\phi_{i}\phi_{i+1}^*+\phi_{i}^*\phi_{i+1})
+\frac{1}{2}\phi_{i}^2{\phi_{i}^*}^2$ is the Hamiltonian of the DNLS 
$i\dot{\phi}_n=\frac{\partial \cal{H}}{\partial \phi_n^*}$. 
Persistent localization of 
energy occurs only if the system's energy is positive. 
\begin{figure}[b]
\epsfbox{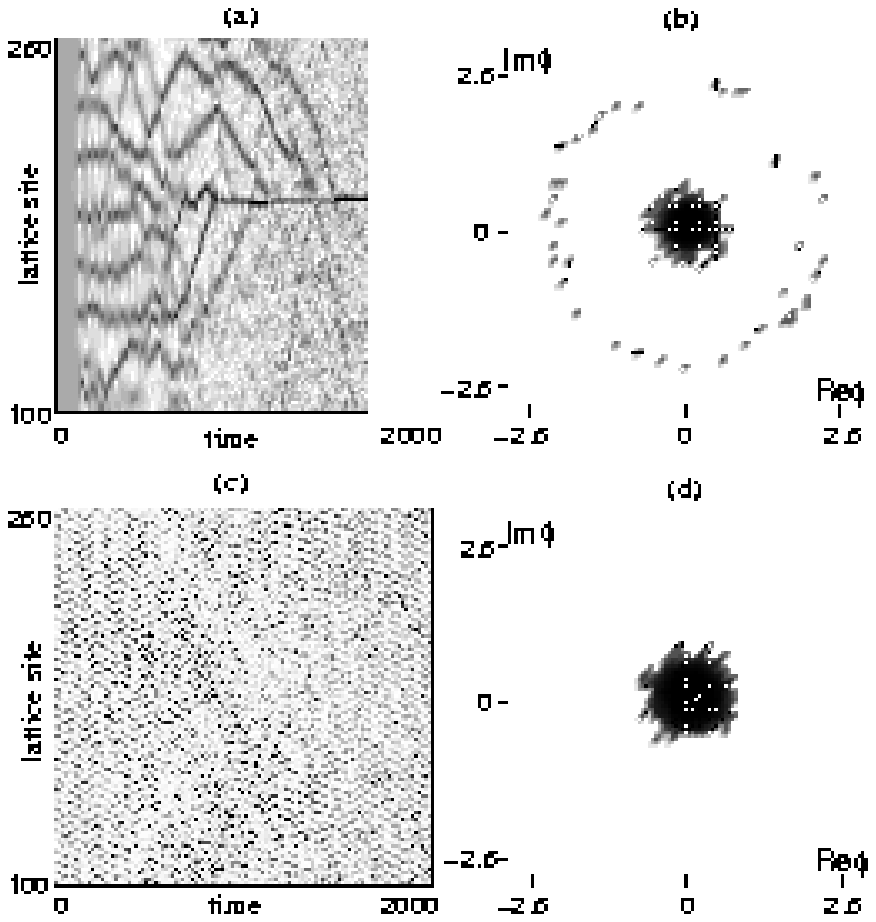}
\caption{
Numerical integration of the DNLS with 4096 oscillators. 
The initial conditions are waves with the wavenumber 
$k=0$ (a),(b), and with $k=\pi/2$ 
(c),(d) for $|\phi|=0.3$. (a),(c): Spatiotemporal pattern of 
high-amplitude states (dark gray) in a small sector of the chain 
for the first 2000 time steps. 
(b),(d): Distribution of $\phi$ after $2\times 10^5$ time steps.  
}
\end{figure}
In this case the peaks finally absorb almost the total energy (Fig.2a). 
The height of the peaks $|\phi|\approx 2...2.5$ is 
almost independent of the total energy (Fig.2b), but the number of peaks 
increases with the energy.  
For negative energies, there is no 
localization of energy (Fig.1c) and the system settles into a state of 
low-amplitude fluctuations (Fig.1d,Fig.2a). 
The system's second conserved quantity, 
the modulus-square norm (or 'particle number') $A=<\cal{A}>$ with 
$
{\cal{A}}=\sum \phi_i\phi_i^*
$
is also crucial for this phenomenon. 
There is no persistent localization of energy 
in this system if the rotational symmetry 
linked to this second conserved quantity is broken. \\
The discrete nonlinear Schr\"odinger equation 
admits energies $-2A\le E \le 2A$ for low-amplitude initial 
conditions $A/N \ll 1$. 
\begin{figure}[t]
\epsfbox{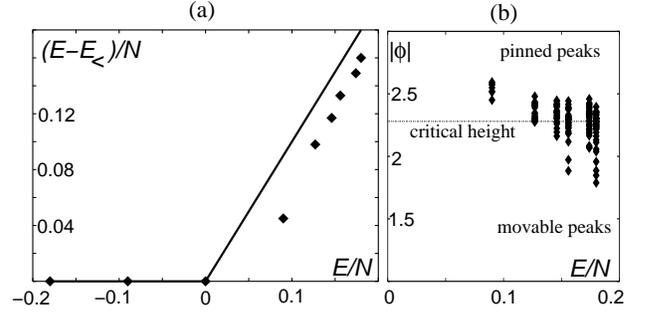}
\caption{
Energy (dots in (a)) and height (dots in (b)) 
of peaks with $|\phi|>1$ as a function of the total energy 
after $2\times 10^5$ time steps of numerical integration. 
The initial conditions of the 4096 lattice sites are 
waves with an amplitude $|\phi|=0.3$ and wavenumbers 
from $k=0$ to $k=\pi$. Lines: Thermodynamic equilibrium value (a) 
and critical height for lattice pinning (b). 
}
\end{figure}
As suggested by Fig.1b, the phase space may be separated 
into a low-amplitude domain 
$\Delta_{<}$ for $|\phi_<|< r $ and a high-amplitude domain 
$\Delta_{>}$ for $|\phi_>|\ge r$. To avoid formal difficulties 
with diverging terms, a preliminary upper bound $R>r$ 
of the phase space is introduced as $|\phi|\le R$. 
The main contributions to the energy arise from interactions 
in the low-amplitude domain (where $N-K$ oscillators are gathered 
and nonlinear contributions are negligible) and from the quartic 
contribution of $K\ll N-K$ oscillators in the high-amplitude domain. 
The Hamiltonian is therefore approximated as 
\begin{equation}
\label{trunchamilton}
\begin{array}{ccl}
{\cal{H}}&\approx&{\cal{H}}_<+{\cal{H}}_>\\
&=&\sum u(r-|\phi_i|)u(|r-|\phi_{i+1}|)
(\phi_{i}\phi_{i+1}^*+\phi_{i}^*\phi_{i+1})\\
&&+\frac{1}{2}u(|\phi_i|-r)\phi_i^2{\phi_i^*}^2
\end{array}
\end{equation}
with the unit step function $u(x<0)=0$, $u(x\ge 0)=1$.
The second integral of motion may be separated in a similar 
way as ${\cal{A}}={\cal{A}}_<+{\cal{A}}_>$. 
The system's entropy as a function of the conserved quantities 
$E$ and $A$ can be computed from the grand 
partition function
$
y(\beta,\gamma)=
\int \rho 
\prod_{i=1}^N
d\phi_id\phi_i^*
$
using the density 
$
\label{density}
\rho=e^{-\beta({\cal{H}}-\gamma \cal{A})}
$. 
The partition function is a sum over all possible numbers 
of peaks $K$, and only its leading term 
\begin{equation}
\label{partition}
y(\beta,\gamma)\approx
{N\choose K}y_<(\beta,\gamma,N-K)y_>(\beta,\gamma,K)
\end{equation}
will be considered. 
$K$ and $R$ will be determined later in order to maximize the entropy. 
The factor ${N\choose K}$ gives the number of combinations 
of $K$ high-amplitude sites on $N$ lattice sites. 
The contributions $y_<$ and $y_>$ will be computed separately. 
The patches of small-amplitude fluctuations between the peaks 
give the contribution  
\begin{equation}
\label{partitionsmall}
y_<(\beta,\gamma)=
\int_{\Delta_{<}}...\int_{\Delta_{<}}
\rho_<  \prod_{i=1}^{N-K} d\phi_id\phi_i^*
\end{equation}
to the partition function with 
$\rho_<=e^{-\beta({\cal{H}}_<-\gamma {\cal{A}}_<)}$. 
These regions are in thermal equilibrium with each other, 
and therefore they are regarded as one single field of 
$N-K$ oscillators. 
Similarly, the $K$ oscillators in 
the high-amplitude regime $\Delta_{>}$ have degrees of freedom 
corresponding to variations of phase and amplitude of the peaks. 
This leads to the contribution to the partition function 
\begin{equation}
\label{partitionbig}
y_>(\beta,\gamma)=
\int_{\Delta_{>}}...\int_{\Delta_{>}}
\rho_>  \prod_{j=1}^K d\phi_jd\phi_j^*
\end{equation}
with 
$\rho_>=e^{-\beta({\cal{H_>}}-\gamma {\cal{A}}_>)}$. 
This neglects the coupling term of the peaks which is small compared 
to the quartic energy. 
However, peaks and fluctuations are coupled  
thermally, so that they can exchange 
energy and particles. This exchange vanishes on average 
when peaks and fluctuations 
have the same 
temperature $\beta^{-1}$ and chemical potential $\gamma$. 
The energy $E\approx E_<+E_>$ and 
the particle number $A=A_<+A_>$ have contributions from 
either domain. $A$, $A_<$, $A_>$, $E_>$ are positive, 
$E$ and $E_<$ may be positive or negative. \\
The partition function $y_<$ can be reduced to Gaussian integrals. 
The density may be written as 
$\rho_<=\prod e^{-(\lambda_1 \phi'_n-\lambda_2 \phi'_{n+1})^2}
e^{-(\lambda_1 \phi''_n-\lambda_2 \phi''_{n+1})^2}$ 
with $\lambda_{1/2}=
(\sqrt{-\beta(\gamma+2)}\pm \sqrt{-\beta(\gamma-2)})/2$ 
(it turns out later that $-\beta\gamma\ge 0$ and $|\gamma|\ge 2$). 
Introducing the variables $x_n=\lambda_1 \phi'_n-\lambda_2 \phi'_{n+1}$, 
the partition function of the fluctuations is reduced to Gaussian integrals  
\begin{equation}
\label{partform}
y_<(\beta,\gamma)=(\int e^{-x_n^2} \det (\partial \phi'_n/\partial x_n)
\prod dx_n)^2
\end{equation}
Using $\det (\partial x_n/\partial \phi'_n)=
\lambda_1^{N-K}\pm\lambda_2^{N-K}$, 
the Jacobian is 
$\det (\partial \phi'_n/\partial x_n)\approx \lambda_1^{-(N-K)}$ 
for $N-K\gg 1$ since $\lambda_1>\lambda_2$. 
The square in equation (\ref{partform}) 
is obtained from the identical integration over $\phi''$. 
This gives an analytic expression for the partition function 
$
y_<(\beta,\gamma,N-K)=\Lambda^{N-K}
$
with 
$\Lambda=\pi\lambda_1^{-2}=\pi(-\beta\gamma+\sqrt{\beta^2(\gamma^2-4)})^{-1}$. 
The corresponding energy is 
\begin{equation}
\label{energysmall}
\begin{array}{ccl}
E_< &=& (\frac{\gamma}{\beta}
\frac{\partial}{\partial\gamma}-\frac{\partial}{\partial\beta})ln y_<\\
&=&(N-K)\frac{\sqrt{\gamma^2-4}-\sqrt{\gamma^2}}{\beta\sqrt{\gamma^2-4}}
\end{array}
\end{equation}
and their particle number is 
\begin{equation}
\label{asmall}
\begin{array}{ccl}
A_< &= &\frac{1}{\beta}\frac{\partial}{\partial\gamma}ln y_<\\
&=&-(N-K)\frac{\gamma}{\beta\sqrt{\gamma^4-4\gamma^2}}
\end{array}
\end{equation}
Consequently, the inverse temperature is 
$\beta=-2E_<(N-K)/(4A_<^2-E_<^2)$ 
and the chemical potential is 
$\gamma=(4A_<^2+E_<^2)/(2E_<A_<)$. 
Using these expressions, the canonic transformation 
$S_<=lny_< +\beta(E_<-\gamma A_<)$ 
leads to the entropy of the small fluctuations 
\begin{equation}
\label{ssmall}
S_<=(N-K)ln \Omega
\end{equation}
with $\Omega=(4A_<^2-E_<^2)/(4A_<(N-K))$. The entropy is the 
most useful thermodynamic potential in this context because 
its variables (particle number and energy) are known 
from the initial conditions.  
The fluctuations entropy per lattice site $ln\Omega$ is 
plotted in Fig.3a. The lines $E_<=2A_<$ and $E_<=-2A_<$ 
correspond to waves with $k=0$ and $k=\pi$ respectively. 
The entropy is infinitely low for these ordered states. 
The ridge $E_<=0$ corresponds to a fluctuating state 
with an infinite temperature where 
all wavenumbers have the same power.
A wave with $k=\pi/2$ 
is a non-thermalized solution with $E=0$. 
This line $\beta=0$ was first computed in 
\cite{ras} where the additional nonlinear 
correction were included, and it was identified as the transition line 
to the localization phase. No analytic results for the statistics 
beyond this line have been available yet. 
\\
Equation (\ref{ssmall}) gives a valid expression for the total 
entropy if the system's total energy is negative (corresponding 
to the right slope with $E_< <0$ of Fig.3a). 
The temperature is positive in this 
case, and consequently the density $\rho_>\sim e^{-\beta\sum |\phi_n|^4/2}$ 
decays rapidly for huge amplitudes 
so that the high-amplitude contribution to the partition function 
is negligible. 
In this phase only small fluctuations contribute to the total entropy 
$S=N ln \Omega$. 
This describes the low-amplitude fluctuating state 
($E=E_<, A=A_<$) with no peaks of Fig.1d. \\
The slope $E_< >0$ is linked to 
negative temperatures \cite{ras}. 
The density $\rho_>$ increases with the amplitude in this regime 
which was suggested to be the reason for the formation 
of high-amplitude structures \cite{ras}. 
The increase of the density causes obvious technical difficulties: 
The phase space is 
(unlike in the spin system of \cite{rune, rune03}) noncompact, 
so that the integral $y_>$ will diverge for any fixed negative 
values of $\beta$. 
However, the artificial upper boundary $R$ of the phase space 
prevents this divergence. The partition function can first be 
computed as a function of this parameter $R$, which subsequently 
will be allowed to go to infinity as this maximizes the entropy. 
From this the contributions of the peaks to the entropy, 
the energy and the particle number can be computed. \\
The grand partition function of the peaks increases as 
$y_>(\beta,\gamma)
\sim(e^{-\beta(R^4/2-\gamma R^2)}/(-\beta R^2+\beta\gamma))^K$ 
with the cut-off $R$ of the phase space if $R$ is greater 
than $\sqrt{\gamma}$. 
Since the density $\rho$ gathers at the border $R$, the particle 
number related to the domain $\Delta_{>}$ is 
$A_>=KR^2$
and the energy is 
$E_>=KR^4/2$. 
Inserting $\beta$, $\gamma$ and $R=\sqrt{2E_>/A_>}$ into the entropy 
of the peaks $S_>\approx -K ln(-\beta R^2)$ yields 
\begin{equation}
\begin{array}{ccl}
S_> & = & K ln(\frac{K(4A_<^2-E_<^2)}{4(N-K)A_>E_<})\\
& = & 
Kln(K/N)+Kln\Omega+Kln\Gamma
\end{array}
\end{equation}
with $\Gamma=2A_<N/A_>E_<$.  
The combinations of $K=A_>^2/2E_>$ peaks 
on $N$ lattice sites 
yield an entropy contribution 
$
S_p=ln {N \choose K} \approx K ln(N/K)
$. 
Adding up $S_<$, $S_>$, and $S_p$, the total entropy is 
\begin{equation}
\label{stotal}
S=N ln \Omega
+\frac{A_>^2}{2E_>} ln\Gamma
\end{equation} 
This expression describes the thermodynamics in the domain 
where localization occurs. 
\begin{figure}[t]
\epsfbox{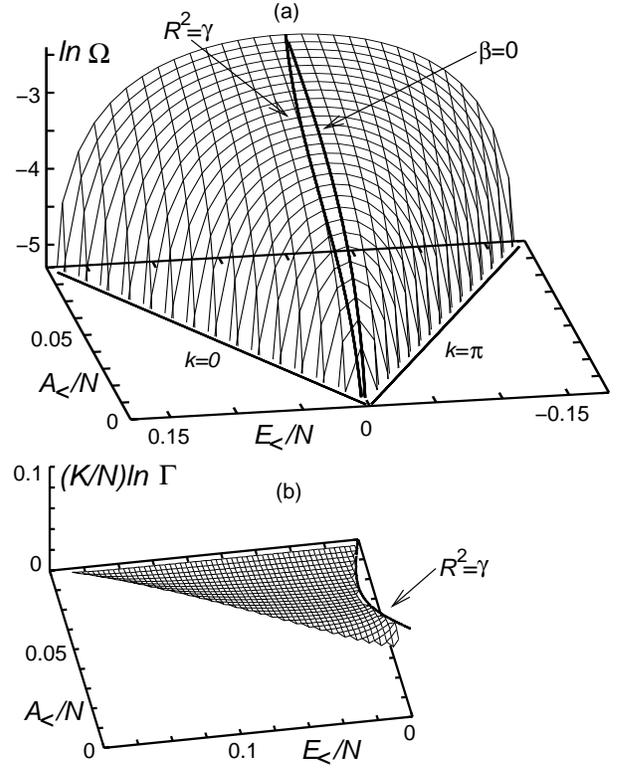}
\caption{
$ln\Omega$ (a) and $(K/N)ln\Gamma$ (b) as a function of $E_<$ and $A_<$; 
line $R^2=\gamma$ for the simulation of Fig.1a,b
}
\end{figure}
The small fluctuations provide the leading term $Nln\Omega$ 
of the entropy while the contribution $K ln\Gamma$ (Fig.3b) 
from the peaks 
is negligible 
under the constraint $R^2>\gamma$ or 
$2E_>/A_> >(4A_<^2+E_<^2)/(2E_<A_<)$. 
$y_>$ and $\Gamma$ can be computed in the same way for 
various types of nonlinearities. 
The entropy increases with $A_<$ and decreases with $|E_<|$ 
and has its maximum at $E_<=0$, $A_<=A$, $E_>=E$, $A_>=0$. 
The state of maximum entropy is related to 
low amplitude fluctuations with $\beta=0$ and 
$\beta\gamma=-N/A$. \\
The crucial point is, that the high peaks contribute little to the 
total entropy, while they can absorb high amounts of energy using 
only few particles. On the other side, the fluctuations can reach 
a state with a maximum entropy, if they contain the ideal amount 
of energy. 
This shows the thermodynamical nature of energy-localization: 
In order to maximize the system's total entropy, the ideal amount 
of energy $E_<$ must be allocated to the fluctuations. 
Starting from an initial condition with a positive energy $E=E_<$ 
at the left slope in Fig.3a, the entropy can be increased when 
$E_<$ decreases while the released energy is stored in the 
localized structures. The state of maximum entropy corresponds to 
only one peak which 
absorbs the total energy (Fig.2a) while consuming very few particles. 
This also shows the 
self-consistency of the truncation (\ref{trunchamilton}) that neglects the 
interaction of the peaks with their environment. 
The equilibrium state with localized structures corresponds to $\beta=0$, 
so the temperature is not negative. 
In the opposite case with 
negative total energies $E$ and a positive temperature, 
there is no energy surplus and consequently 
no localization (Fig.1c,d). 
\\
Fig.4 gives a numerical picture of the shift of energy and particles 
into the peaks for equation (\ref{dnls}) and the corresponding picture 
for a version of the DNLS where the 
rotational symmetry is broken by a small term $0.015 Re(\phi_n)$. 
It shows the evolution of the 
fluctuations energy $E_<$ and particle number $A_<$ 
versus the corresponding equilibrium isentropes $S=const$ for 
a homogeneous low-amplitude ($|\phi|=0.3$) initial state 
$E_<=E=2A$ and $A_<=A$. 
\begin{figure}[t]
\epsfbox{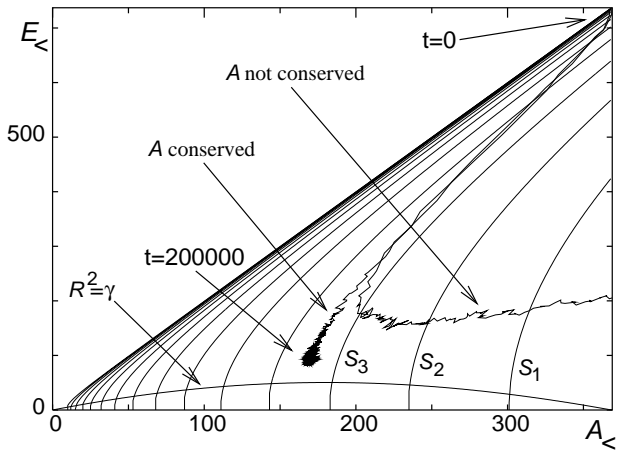}
\caption{
Isentropes $S_1>S_2>S_3>...$ for the left slope of Fig.3a; 
trace $(E_<,A_<)$ of the integration of Fig.1a,b over $E_<$, $A_<$; 
trace for a version of the DNLS where the second 
conservation law is violated.
}
\end{figure}
The boundary between $\Delta_<$ and $\Delta_>$ is $r=1$. 
The particle-nonconserving system produces no persistent peaks. 
Instead, it increases its entropy by increasing the total number 
of particles and ends up in a fluctuating 
high-amplitude state \cite{krum}. 
The particle-conserving system (\ref{dnls}) approaches the 
entropy-maximum by generating peaks and transferring energy from the 
fluctuations $E_<$ to 
$E_>$. On the other side, growing peaks also 
absorb particles that must be shifted from $A_<$ to $A_>$. 
The trace $(E_<,A_<)$ therefore strives down to the left in Fig.4. 
This loss of particles in the fluctuations is unfavorable for the 
production of entropy. 
The merging of peaks however allows the system to store 
more energy in the peaks while feeding particles 
from the peaks back into the fluctuations 
$A_<$ which leads to an additional increase of the entropy. 
\\
A state with only one huge peak is not reached experimentally (Fig.1b). 
Instead, a number of coherent structures of moderate 
height (Fig.2b) survive even on very long time scales ($10^7$ time steps in
numerical simulations). 
In Fig.4, the trace $(E_<,A_<)$ does not reach the state of 
maximum entropy at $(E_<=0,A_<\approx 370)$, but ends up at in a state 
with only half that particle content, and still a positive amount of energy. 
The coherent structures contain a significant amount of particles, 
and the entropy of the fluctuations is below its maximum. 
The reason for this is the eventual breakdown of the two 
entropy-enhancing mechanisms: Firstly, the merging of peaks becomes 
impossible because peaks above a certain critical height are immobilized by  
a lattice-pinning effect. For that reason the trace $(E_<,A_<)$ in 
Fig.4 does not cross the isentropes $S_3, S_2, S_1$. 
Secondly, the growth of peaks 
stops when this leads to no further increase of the fluctuations 
entropy. This happens when the entropy gain due to the energy transfer to the 
peaks is matched by the entropy loss due to the particle transfer. 
For this reason, the trace $(E_<,A_<)$ in Fig.4 does not approach the line 
$E_<=0$ any further after $\sim 10^5$ time steps. 
\\
The mobility of a peak with 
an amplitude $|\phi_n|=\sqrt{a}$ at a site $n$  
depends on the compatibility of the local conservation of its energy 
$E=a^2/2$ and its particle number $A=a$ 
during a possible migration to the site $n+1$. 
Most of the $a$ particles are gathered at these two 
lattice sites during the migration. 
The trajectory of such a migration is therefore 
close to the intersection of the level set of $E$ and $A$ with 
the constraint of 
nonvanishing amplitudes at $n$ and $n+1$ only. 
These paths are given by 
$\phi_n(\nu)=\sqrt{a}\sqrt{1-\nu}e^{i\psi}$ and 
$\phi_{n+1}(\nu)=\sqrt{a\nu} e^{i\alpha (\nu)}e^{i\psi}$ 
with $\cos \alpha =a\sqrt{\nu-\nu^2}/2$. The parameter $\nu$ 
goes from $0$ to $1$ during the migration as the particles 
are shifted from $n$ to $n+1$. $\psi$ is a phase factor. 
The bottleneck of this process is the intermediate storage of 
$|\phi|^4$-energy in the coupling term 
$2Re(\phi_n\phi_{n+1}^*)$: 
The solvability condition $\cos \alpha\le 1$ 
for all $\nu$ requires that $a\le 4$, so that this 
migrational path only exists for peaks below a maximum 
amplitude $|\phi_n(\nu=0)|\le 2$. 
For higher peaks, it is impossible to conserve both 
particle number and energy at the instant when 
$|\phi_n|=|\phi_{n+1}|$. 
These idealized migrational paths deviate from exact solutions 
of the DNLS which necessarily have nonvanishing amplitudes
for adjacent oscillators at $n-1$, $n+2$, etc. 
Monte-Carlo simulations of paths that include 
small amplitudes for adjacent oscillators 
(4-20 lattice sites) show that migrations for  
slightly higher peaks ($|\phi_n(\nu=0)|< 2.28$, Fig.2b) are permitted 
by the conservation laws.  
Higher peaks first need to decrease by transferring 
particles to remote lattice sites before they can move. 
This defocusing process would require a very 
unlikely decrease of entropy and 
the local conservation of particles and energy is therefore 
statistically favorable. Spontaneous collapses of peaks are 
only possible if this increases the total entropy, which 
is the case if the fluctuations have small wavelengths and a 
positive temperature.\\
Further growth of the peaks 
requires a flow both of particles and of energy from the 
fluctuations to the peaks as the number of pinned peaks is fixed on relevant 
time scales. This is thermodynamically favorable only 
if the increase of the entropy due to the energy transfer is bigger 
than the decrease caused by the particle transfer. This process 
stops when the trace $(E_<,A_<)$ in Fig.4 approaches an isentrope 
tangentially as  
$\partial E_>/\partial A_>|_{K=const}=\partial E_</\partial A_<|_{S=const}$ 
which is equivalent to $\gamma=R^2$. 
On this line, growth or decay processes of peaks absorb or release 
energy and particles in a ratio that amounts to isentropic 
changes of the fluctuations. 
The statistical results reflect microscopic dynamical 
findings \cite{joha} of the growth and decay processes of localized structures 
perturbed by one or two incoming waves. Long waves lead to growth, 
short waves to decay of a peak while the radiated harmonics increase 
the systems entropy. 
Experimentally, one still finds 
irregular oscillations of the peak heights, but no average 
growth. Interactions of peaks and fluctuations that increase the peaks 
are matched on average by those interactions that 
decrease the peaks. 
Waves with all wavenumbers 
coexist in the state of equilibrium 
where growth processes are balanced by decay processes.  
Any changes of the peak amplitude are statistically 
unfavorable as they decrease the total entropy. 
\\
To conclude, localization 
in non-integrable systems constrained by two integrals of motion 
is a {\itshape statistical\/} process. 
The entropy is dominated by contributions from small-amplitude 
fluctuations. The entropy is maximal if an optimal  
share of each conserved quantity is allocated to the fluctuations. 
There is no localization if not enough energy is 
supplied by the initial conditions. 
If a surplus of energy is provided by the initial 
conditions, it is dumped into high-amplitude structures 
that absorb high amounts of energy (Fig.2) while using 
few particles. 
This explains the phase where energy-localization occurs. 
The lattice pinning effect related to the conservation laws prevents 
the system from reaching the absolute entropy maximum. For the 
final peak size, growth- and decay-interactions of the peaks 
with the fluctuations are balanced. \\
Obviously, 
the statistical analysis shows the macroscopic properties of 
an ensemble of microstates, and one might expect that 
the Arnold diffusion process transfers the trajectory from any 
initial condition to this most probable state. However, 
the thermalization may have extremely long transients. 
In the numerical simulations, the amplitudes in the 
initial conditions were small enough so that 
the separation of the partition 
function in the statistical description was valid, 
but large enough for the nonintegrability 
to have an impact on moderate time scales. 
In the 'integrable' limit of 
smooth initial conditions with very low amplitudes, 
the dynamics is most similar to the continuous onedimensional 
nonlinear Schr\"odinger equation, which is integrable. Consequently, 
this part of the phase space is partitioned by Kolmogorov-Arnold-Moser 
tori that are not destroyed by the nonintegrability and the 
dynamics is quasiperiodic on very long time scales. 
To escape from this state, the amplitude has to 
reach a sufficient height so that the nonlinearity 
absorbs a substantial amount of energy, which becomes a rare event 
for small average amplitudes. 
Such quantities that are almost conserved on moderate time scales can  
also be relevant for localization effects in systems 
where the Hamiltonian is the only exactly conserved quantity. 
For instance, the DNLS with broken rotational symmetry shows 
energy localization on shorter time scales where the particle number 
changes very little. This is the case in the early stadium 
of the path $E_<,A_<$ where $A$ is not conserved in Fig.4. 
It is an interesting question if 
this also applies to other breather systems where the Hamiltonian 
is the only conserved quantity \cite{mojo,dape}. 
\\
The mechanism of localization is found in very diverse dynamical systems 
as it relies only on the properties of the entropy functional and on the 
existence of two conserved quantities and not on the spatial 
discreteness. Depending on the type of the nonlinearity, the 
localized structures may absorb primarily the second conserved 
quantity \cite{rune} and not the energy. 
In continuous systems 
described by partial differential equations 
\cite{zss,dya}, the transport of 
fluctuations is continued down to the molecular scale. 
Again, the the conservation laws require the 
formation of localized structures for this  
exploitation of degrees of freedom on short 
space scales.

\end{document}